# Breaking the width-scaling limit in high-performance atomically thin 2D nanoribbon transistors

Sameer Kumar Mallik[1], Adrian Christiansen[1], Saroj P. Dash[1*]

[1]*Department of Microtechnology and Nanoscience, Chalmers University of Technology, SE-41296, Gothenburg, Sweden.*

## Abstract

**State-of-the-art transistors have been successfully scaled the gate lengths and channel thicknesses down to 5 nm for high-performance and energy-efficient information processing. However, reducing channel width below 40-50 nm remains a bottleneck, as dangling bonds, edge disorder, and lateral depletion suppress drive current and degrade device performance. Here, we break this width-scaling wall using ultra-scaled two-dimensional semiconductor (2DSC) nanoribbon transistors down to 15 nm. In contrast to the conventional scaling rule of degradation of current density upon width scaling, our atomically-thin monolayer and bilayer molybdenum disulfide nanoribbon transistors exhibit enhancement of on-current density of up to 230% and 170%,respectively, followed by a saturation for the narrowest channels down to 15 nm. The ultra-narrow nanoribbon transistors maintain the highest on/off ratios reported so far for similar device dimensions, with improved mobility and threshold-voltage stability, indicating reduced edge scattering and depletion with a stronger electrostatic control. These findings lead to a breakthrough in width scaling rules using 2DSC nanoribbons with enhanced performance at narrower channel widths, which is promising for the ultimate scaling of transistors.**

**Corresponding Author:** saroj.dash@chalmers.se

## Introduction

Over 75 years, transistors have shaped modern nanoelectronics by progressive down-scaling, which has delivered exponential improvements in computational speed, energy efficiency, and integration density[1]. Silicon complementary metal-oxide-semiconductor (CMOS) field-effect transistors (FET) technology has driven the information revolution through progressive scaling[2]. To sustain Moore's-law, transistor architectures have evolved from planar MOSFETs to FinFETs, and more recently to gate-all-around (GAA) nanosheet transistors[3]. Although gate lengths and channel thicknesses have been scaled down to the few-nanometer regime[4], the channel-width scaling in state-of-the-art transistors has saturated at ~40-50 nm (Fig. 1a)[5,6]. Furthermore, p-type MOSFETs require channels that are two or three times wider than n-type devices to compensate for drive current due to lower hole mobility, thereby limiting further CMOS scaling and integration density[7]. Together, these constraints give rise to a "width-scaling wall".

To address these issues, atomically-thin two-dimensional semiconductors (2DSC) such as molybdenum disulfide ($MoS_2$) have emerged as leading candidates for next-generation transistor channels due to their sizable bandgap, balanced electron-hole mobility, and high drive currents [8–12]. Their atomically thin structure provides immunity to short-channel effects, suppresses vertical electrostatic leakage, and makes them natural candidates for GAA nanosheet devices[13–16]. However, most high-performance 2D transistor demonstrations still rely on micrometer-wide channels[17–19], whereas narrower nanoribbons have exhibited degraded electrical performance[20–22]. It is commonly assumed that reducing channel width introduces pronounced edge-roughness induced scattering and lateral depletion at the edges (Fig. 1b), which degrade drive current and reduce device performance parameters such as mobility and sub-threshold swing[21–24], making a width scaling wall in semiconductor technology.

Here, we demonstrate that this "width scaling wall" can be overcome by observing enhanced performance of 2DSC $MoS_2$ monolayer and bilayer nanoribbon transistors down to 15 nm width. Remarkably, by reducing the channel width to 30 nm, we observed an increase in the normalized on-current density and saturation for narrower ribbons, revealing an edge-enhanced transport regime driven by reduced scattering and stronger electrostatic control at the atomically thin nanoribbon edges. This observation is in contrast with the current degradation typically observed in conventional semiconductors with width scaling down to sub-50 nm (Fig. 1b). Notably, our $MoS_2$ nanoribbon transistors show enhancement in device parameters such as on-current density, mobility and threshold-voltages with width down-scaling. By correlating nanoscale structural characterization with electrical transport, our results establish that nano structuring techniques for electrostatically engineered $MoS_2$ nanoribbons could be promising architectures for ultimate width scaling of transistor technologies beyond the present state-of-the-art.

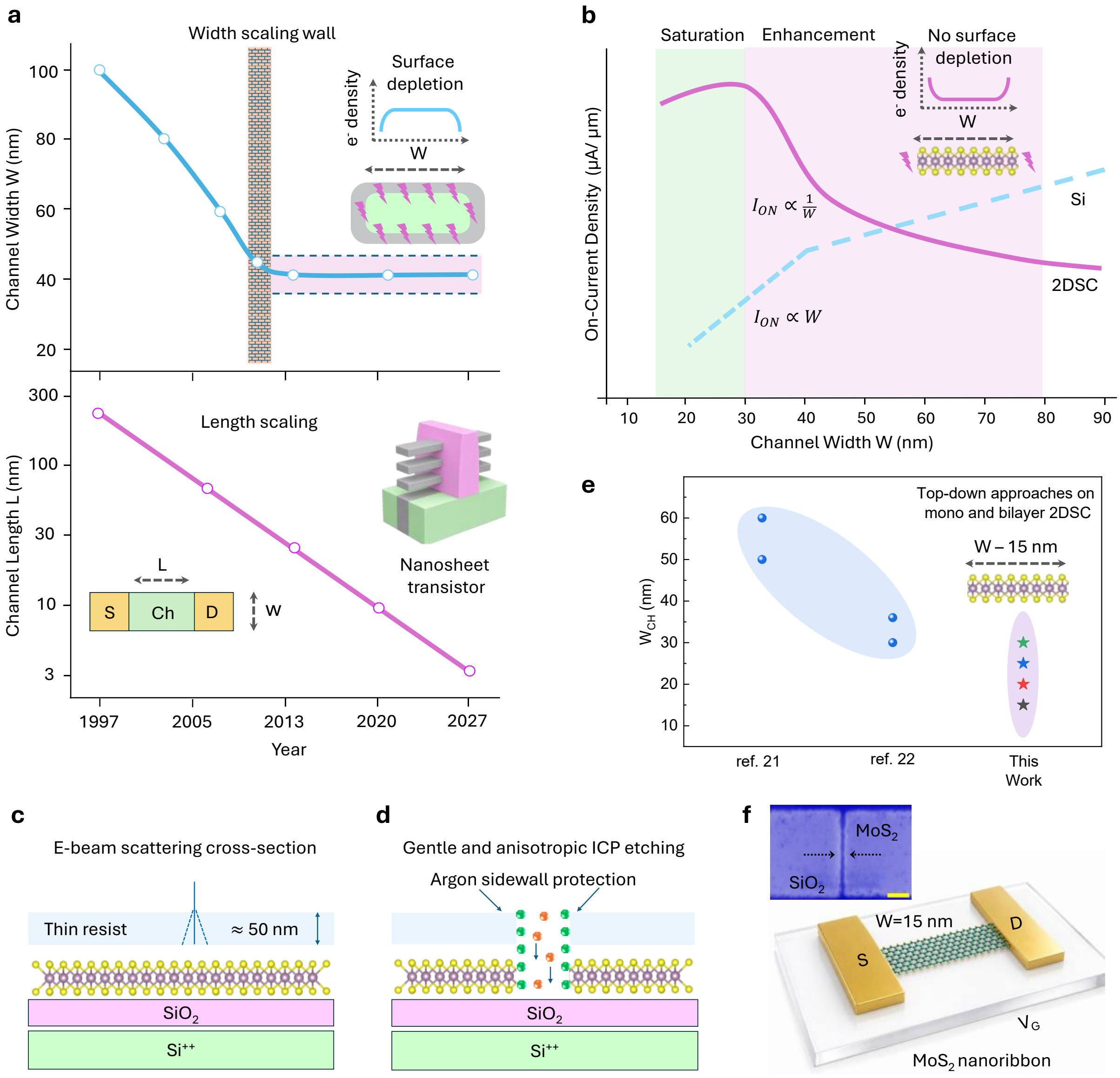

**Figure 1. Width scaling wall in advanced transistors and opportunity for 2D semiconductor nanoribbons. (a)** Historical technology trends for channel width $W$ (top) and channel length $L$ (bottom). Nanosheet transistors show continuous $L$ scaling down to a few nanometres, whereas $W$ has saturated around ~40-50 nm showing a practical "width-scaling wall" due to all-surface scattering at narrow widths. **(b)** Conceptual schematics of on-current density $I_{\mathrm{ON}}$ versus channel width $W$ for Si nanoribbons (cyan) and 2DSC nanoribbons (magenta) highlighting an edge-enhanced regime (shaded) in which only lateral edge states contribute to scattering in 2DSC. **(c)** Schematic of electron-beam exposure with a thin resist stack (~50 nm) used to define $MoS_2$ nanoribbons. **(d)** Schematic of the inductively coupled plasma (ICP) etching with argon-assisted sidewall protection resulting sharp nanoribbon sidewalls. **(e)** Comparison of channel widths of our $MoS_2$ nanoribbons with the state-of-the-art top-down nanoribbon fabrication processes on mono and bilayers. **(f)** False-colour SEM image and three-dimensional schematic of a back-gated $MoS_2$ nanoribbon transistor with a 15 nm-wide channel; Scale bar is 100 nm.

## Results

### Fabrication of high-quality atomically thin monolayer and bilayer $MoS_2$ nanoribbons

Defining atomically sharp nanoribbons with controlled widths approaching the few-nanometer regime, while minimizing the process-induced delamination and unintentional edge doping, remains a major nanofabrication challenge[25–27]. To address this, we develop an optimized top-down fabrication process to maximize the yield of narrower mono and bilayer $MoS_2$ nanoribbons. As illustrated in Fig. 1c-d and Supplementary Fig. S1, two process parameters are critical for reliable sub-50 nm patterning; First, the electron-beam resist thickness is reduced to ∼50-60 nm, minimizing forward scattering, and hence the lateral spread of secondary electrons. This also allows a balanced height-to-width aspect ratio of the exposed nanopatterned area, as confirmed by atomic force microscopy characterizations (see Supplementary Fig. S2). Second, anisotropic inductively coupled plasma etching, assisted by argon-mediated sidewall protection, suppresses lateral undercut and preserves edge fidelity[28,29]. Detailed fabrication procedures are provided in Methods and Supplementary Note S1-S2. This process enables statistical sampling across atomically-thin nanoribbons down to 30 nm widths, while retaining the scalability advantages over catalyst-growth-based nanoribbon synthesis routes[30–36]. Figure 1e shows a comparison of our nanofabricated $MoS_2$ nanoribbon channel width with the state-of-the-art results with transport measurements using top-down fabrication processes on atomically thin mono and bilayer 2DSCs, where channel width is limited by the resolution of lithography and physical etching techniques. False-colour scanning electron microscopy (SEM) image of a representative monolayer $MoS_2$ nanoribbon with a channel width of ∼15 nm is presented in Fig. 1f. Supplementary Fig. S3 highlights SEM images of fabricated nanoribbons for wider ranges up to 200 nm, demonstrating precise control over the nanoribbon width across nearly an order of magnitude.

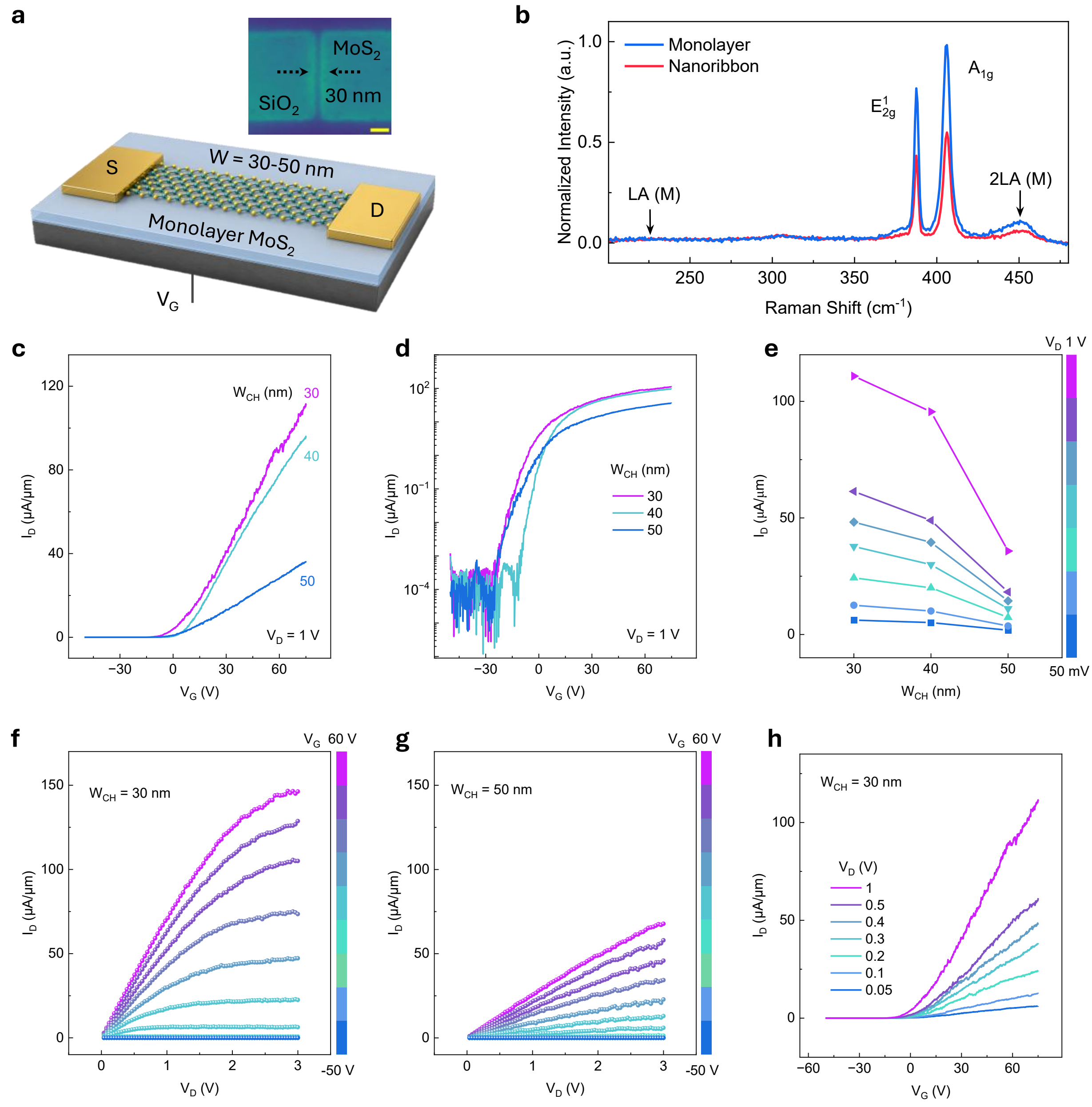

**Figure 2. Enhancement of Electrical transport properties of monolayer $MoS_2$ nanoribbon transistors as a function of channel width down to 30 nm. (a)** False-colour SEM image and schematic of a back-gated monolayer $MoS_2$ nanoribbon transistor with a 30 nm-wide channel; Scale bar is 100 nm. **(b)** Raman spectra of monolayer $MoS_2$ (blue) and patterned $MoS_2$ nanoribbons (red), showing dominant vibrational modes. **(c)** Linear-scale transfer characteristics ($I_D$ vs. $V_G$) for $MoS_2$ nanoribbon transistors with channel widths $W_{CH}$ = 30, 40, and 50 nm at $V_D$ = 1 V and $L_{CH}$ = 50 nm. **(d)** Corresponding semi-logarithmic transfer characteristics for the same devices demonstrate n-type transistor behavior with a high on/off current ratio. **(e)** On-current density ($I_D$) as a function of $W_{CH}$ for drain voltages ranging from $V_D$ = 50 mV to 1 V, demonstrating a clear enhancement of on-current with decreasing channel width across all bias conditions. **(f, g)** Output characteristics ($I_D$ vs. $V_D$) for devices with $W_{CH}$ = 30 nm and 50 nm, respectively ($L_{CH}$ = 50 nm), measured at gate voltages swept from $V_G$ = −50 V to 60 V. **(h)** Transfer characteristics of the $W_{CH}$ = 30 nm, $L_{CH}$ = 50 nm device at multiple drain biases ($V_D$ = 50 mV to 1 V).

**Enhanced regime electrical transport in mono and bilayer $MoS_2$ nanoribbon transistors**

The monolayer $MoS_2$ nanoribbon transistors with channel widths from 50 to 30 nm are fabricated on $SiO_2$ (285 nm)/$Si^{++}$ substrates, which serve as back gates with Ni/Au source and drain contacts, as shown schematically in Fig 2a. To evaluate the structural integrity of the fabricated nanoribbons, the Raman spectroscopy is performed on the nanoribbons and compared with pristine monolayer $MoS_2$ (Fig. 2b). The characteristic $E_{2g}^1$and $A_{1g}$ phonon modes exhibit no measurable peak shift or linewidth broadening following nanoribbon patterning (Supplementary Fig. S4)[37]. In addition, the defect-activated LA(M) mode near ∼227 $cm^{-1}$, which is typically enhanced in plasma-damaged or disordered $MoS_2$[38], remains negligible. The nearly identical Raman signatures between pristine and patterned regions indicate that the fabrication process preserves the crystallinity and edge quality of the $MoS_2$ nanoribbons. The linear-scale transfer curves in Fig. 2c show that, at a fixed channel length of $L_{\mathrm{CH}} = 50$ nm and drain bias $V_{\mathrm{D}} =$ 1 V, the on-current density systematically increases from ∼35 to 110 µA/µm as the channel width is reduced from 50 to 30 nm, respectively. This behaviour contrasts with earlier monolayer $MoS_2$ nanoribbon FETs fabricated by scanning-probe or plasma etching, where trimming the channel width from 60 to 30 nm generally led to reduced mobility, lower on/off ratio and enlarged subthreshold swing due to edge disorder and process damage[21,24]. The semi-logarithmic plots in Fig. 2d confirm n-type conduction with steep turn-on and low off-currents maintaining high on/off ratio of $\sim10^6$ for all widths, indicating that narrowing the ribbon does not introduce excessive leakage channels, unlike some previous monolayer and multilayer nanoribbon studies in which nanoribbon formation created additional series barriers between micro- and nano-segments of the channel[22,24]. The on-current density versus $W_{\mathrm{CH}}$ in Fig. 2e reveals a clear monotonic enhancement of $I_{\mathrm{D}}$ as the channel is narrowed, and this trend persists from near-ohmic bias at $V_{\mathrm{D}} = 50$ mV (see Supplementary Fig. S5) up to $V_{\mathrm{D}} = 1$ V. Such a consistent width dependence at both low and high fields is distinct from multilayer $MoS_2$ nanoribbon and graphene nanoribbon FETs, where size effects and edge roughness typically suppress the current at the narrowest widths[24,25]. Our data therefore support a picture of "electrostatic edge enhancement" in atomically thin nanoribbons, in which the gate field couples more efficiently into the channel as the lateral dimension approaches a few tens of nanometers, overcoming the increase in edge scattering, and carrier depletion that limits performance in thicker nanoribbon devices.

The output characteristics in Fig. 2f, g further highlight the role of narrow nanoribbons for current injection and saturation. For $W_{\mathrm{CH}} = 30$ nm, the $I_{\mathrm{D}}$-$V_{\mathrm{D}}$ curves show a high saturation current density and an early onset of current saturation as the gate voltage is increased from -50 to 60 V, indicating strong gate control over the entire ribbon and efficient modulation of the channel charge. By contrast, the 50 nm-wide device (Fig. 2g) reaches lower saturation currents under the same biases due to a weaker lateral gate coupling. The absence of pronounced kinks or negative differential conductance in these output curves suggests that thermal effects and contact-limited transport are not yet dominant in this bias window[39,40]. Furthermore, the transfer characteristics of the $W_{\mathrm{CH}} = 30$ nm device at multiple drain biases (Fig. 2h) demonstrate robust gate modulation, a high on/off ratio ($\sim 10^6$) and a systematic increase of on-current with $V_{\mathrm{D}}$ (see Supplementary Fig. S6 for $W_{\mathrm{CH}} = 40$ nm and $50$ nm devices). The equivalent gate-oxide field at $V_G$ = 60 V on 285 nm $SiO_2$ corresponds to $\sim$2.1 MV/cm, well within the operating range of thin high-K gate stacks (1-3 MV/cm). These measurements therefore reflect an intrinsic property of the $MoS_2$ nanoribbon channel and integration with thin $HfO_2$ gate stack is expected to achieve the same performance at $V_G$ below 2 V, relevant for low-voltage operation.

Bilayer $MoS_2$ represents a promising channel material for advanced transistor scaling compared to monolayer channels with higher carrier density, lower contact resistance and improved mobility. Motivated by these advantages, we further investigate the electrical transport studies of bilayer $MoS_2$ nanoribbons for four different channel widths in 30-80 nm range, as schematically illustrated in Fig 3a. Normalized Raman spectra of bilayer $MoS_2$ and a lithographically defined nanoribbon show the characteristic first-order $E_{2g}^{1}$ and $A_{1g}$ modes together with the low-frequency LA(M) feature (Fig. 3b). The nearly identical linewidths with negligible peak shifts demonstrate that nanoribbon patterning preserves the vibrational fingerprints of bilayer $MoS_2$ (Supplementary Fig. S7). At fixed channel length $\mathrm{L}_{\mathrm{CH}} = 200$ nm and drain bias $\mathrm{V}_{\mathrm{D}} = 2$ V, the linear-scale transfer characteristics of bilayer $MoS_2$ nanoribbon transistors exhibit a systematic increase in on-current density from $\sim$125 to 300 μA/μm as the channel width is reduced from 80 to 30 nm as shown in Fig. 3c. In contrast to earlier studies on thick $MoS_2$ nanoribbon transistors, where reducing the channel width below 100 nm typically caused mobility and on-current degradation due to edge roughness and defect scattering[20], our atomically thin $MoS_2$ nanoribbons exhibit enhanced current transport at narrower widths. The semi-logarithmic transfer curves in Fig. 3d show high on/off ratios $\sim 10^6$ and steep turn-on for all widths, indicating that bilayer nanoribbon does not introduce dimensional-scaling effects such as excessive leakage paths or strong parasitic barriers[41]. These results on bilayer nanoribbons suggest that carrier depletion at edges is minimized and the electrostatic benefit of lateral confinement can be harnessed without sacrificing channel integrity, in contrast to recent demonstrations of edge-engineered $WSe_2$ nanoribbon transistors, where edge passivation improved the electrical transport[42]. The on-current density versus channel width shows a monotonic enhancement of $I_{\mathrm{D}}$ as $W_{\mathrm{CH}}$ is reduced from 80 nm to 30 nm (Fig. 3e), and this trend persists from low bias ($V_{\mathrm{D}} = 0.5$ V) up to $V_{\mathrm{D}} = 3$ V.

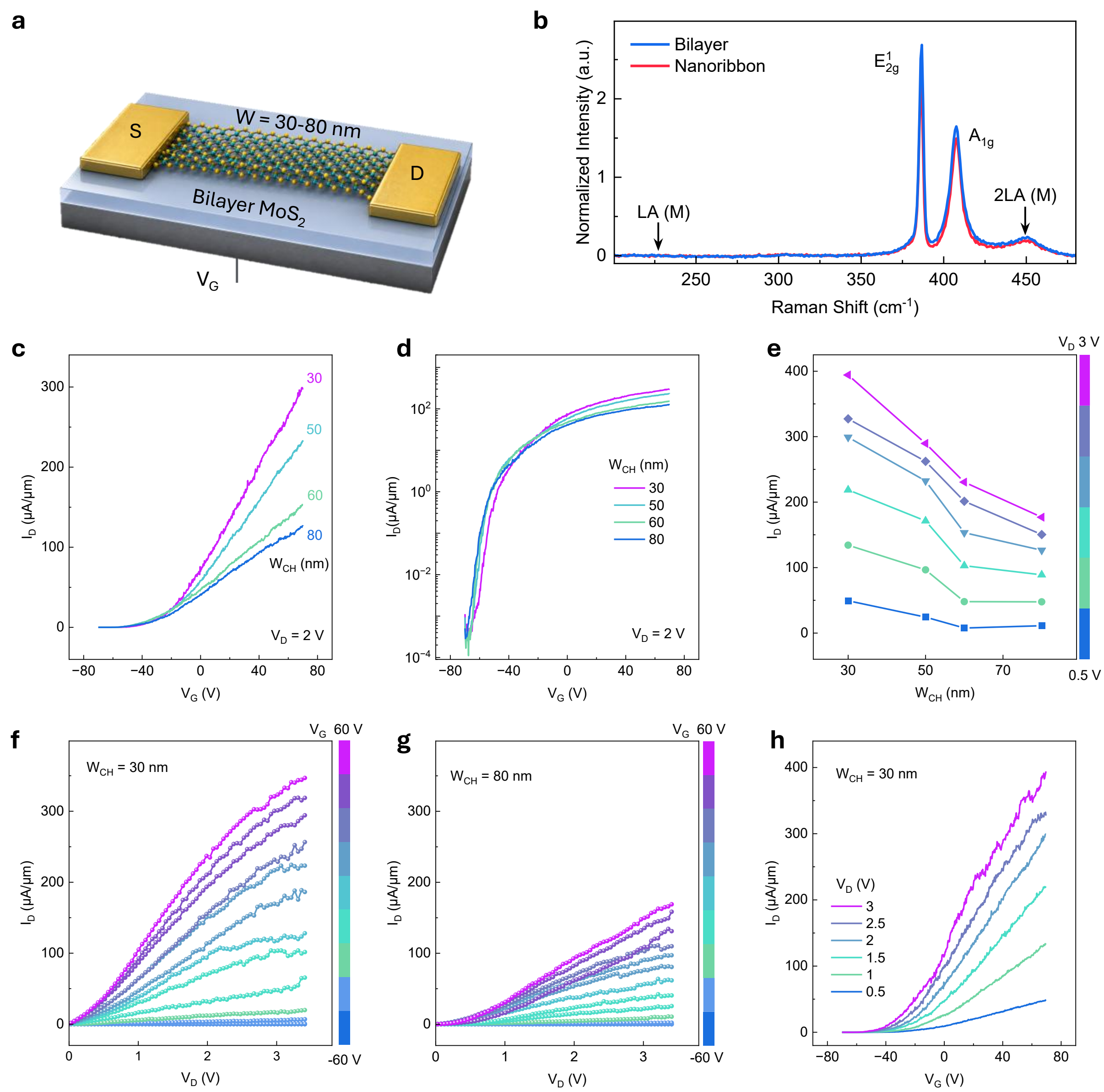

**Figure 3. Enhancement of Electrical transport properties of bilayer $MoS_2$ nanoribbon transistors as a function of channel width down to 30 nm. (a)** Schematic of a back-gated bilayer $MoS_2$ nanoribbon transistor. **(b)** Raman spectra of bilayer $MoS_2$ (blue) and patterned $MoS_2$ nanoribbons (red), showing dominant vibrational modes. **(c)** Linear-scale transfer characteristics ($I_D$ vs. $V_G$) for bilayer $MoS_2$ nanoribbon transistors with channel widths $W_{CH}$ = 30, 50, 60, and 80 nm at $V_D$ = 2 V and $L_{CH}$ = 200 nm. **(d)** Corresponding semi-logarithmic transfer characteristics for the same devices, demonstrating n-type transistor behavior with a high on/off current ratio. **(e)** On-current density ($I_D$) as a function of $W_{CH}$ for drain voltages ranging from $V_D$ = 0.5 V to 3 V, showing a monotonic enhancement of on-current with decreasing channel width across all bias conditions. **(f, g)** Output characteristics ($I_D$ vs. $V_D$) for devices with $W_{CH}$ = 30 nm and 80 nm, respectively ($L_{CH}$ = 200 nm), measured at gate voltages swept from $V_G$ = −60 V to 60 V. **(h)** Transfer characteristics of the $W_{CH}$ = 30 nm, $L_{CH}$ = 200 nm device at multiple drain biases ($V_D$ = 0.5-3 V), showing a significant on-current enhancement reaching ~400 μA/μm at $V_G$ = 70 V.

The output characteristics in Fig. 3f, g further highlight the advantage of narrow bilayer nanoribbons for current injection and saturation. For $W_{CH}$ = 30 nm, the $I_D$-$V_D$ curves exhibit a high saturation current density and an early onset of current saturation as the gate voltage is increased from -60 V to 60 V, indicating strong gate control and efficient modulation of the channel charge. In contrast, the 80 nm-wide device reaches lower saturation currents under the same bias conditions and exhibits more gradual output characteristics, consistent with weaker lateral gate coupling and a larger effective electrostatic length. Figure 3h shows the bias-dependent transfer curves of the 30 nm wide device measured at different $V_{\mathrm{D}} = 0.5$-3 V, which displays robust gate modulation and steadily increasing on-current with drain bias (see Supplementary Fig. S8 for 50, 60, and 80 nm wide devices). The on-current density approaches $400\ \mu A/\mu m$ at $V_G = 70\ V$ and drain bias $V_D = 3\ V$, which is comparable to or higher than the drive currents of state-of-the-art bilayer and few-layer $MoS_2$ transistors with $L_{\mathrm{CH}}$ = 35 nm without nanoribbon confinements[18,43].

The enhanced performance of 2D semiconductor nanoribbons up to 30 nm channel widths can be attributed to the following mechanisms. First, edge-induced electric field enhancement leads to increased carrier accumulation near the ribbon boundaries, improving gate electrostatic control in the on-state. Second, edge-related doping, including passivation of sulfur vacancies, can suppress mid-gap states and reduce contact resistance. Third, edge scattering appears to play a negligible role at these dimensions, as the electron mean free path in monolayer $MoS_2$ (~5 nm) is significantly shorter than the nanoribbon width. Collectively, these effects suggest the potential scaling of 2DSC nanoribbons beyond the existing state-of-the-art silicon nanoribbon limits.

## Ultimate-width scaling down to 15 nm of $MoS_2$ nanoribbon transistors

With careful optimization of the nanofabrication process, the 2DSC channel width is further reduced to an ultra-narrow regime, down to 15 nm (Fig. 4a), and thereby demonstrating high-performance bilayer $MoS_2$ nanoribbon transistors. False-colour SEM images and corresponding transfer characteristics of representative ultra-narrow bilayer $MoS_2$ nanoribbon transistors with channel widths of 25, 20 and 15 nm are shown in Fig. 4b,c, respectively. We observe that the on-current density levels off and saturates at around 200 μA/μm at $V_D$ = 2 V. The semi-logarithmic transfer curves in Fig. 4d exhibit nearly overlapping threshold voltages and steep subthreshold characteristics across all widths, indicating that extreme width scaling does not introduce much performance degradation. Instead, the transport behaviour suggests a transition into a saturation regime governed by the balance between enhanced edge electrostatics and the onset of edge-related scattering. Consistently, the width-dependent on-current densities extracted at drain biases from 0.5 to 3 V (Fig. 4e) confirm that the edge-enhanced transport regime identified between 80 and 30 nm gradually evolves into a saturation regime as the channel width approaches the ultimate scaling limit of 15 nm. These observations establish a new dimensional scaling rule for atomically thin $MoS_2$ nanoribbons, beyond which electrostatic gains are independent of channel widths.

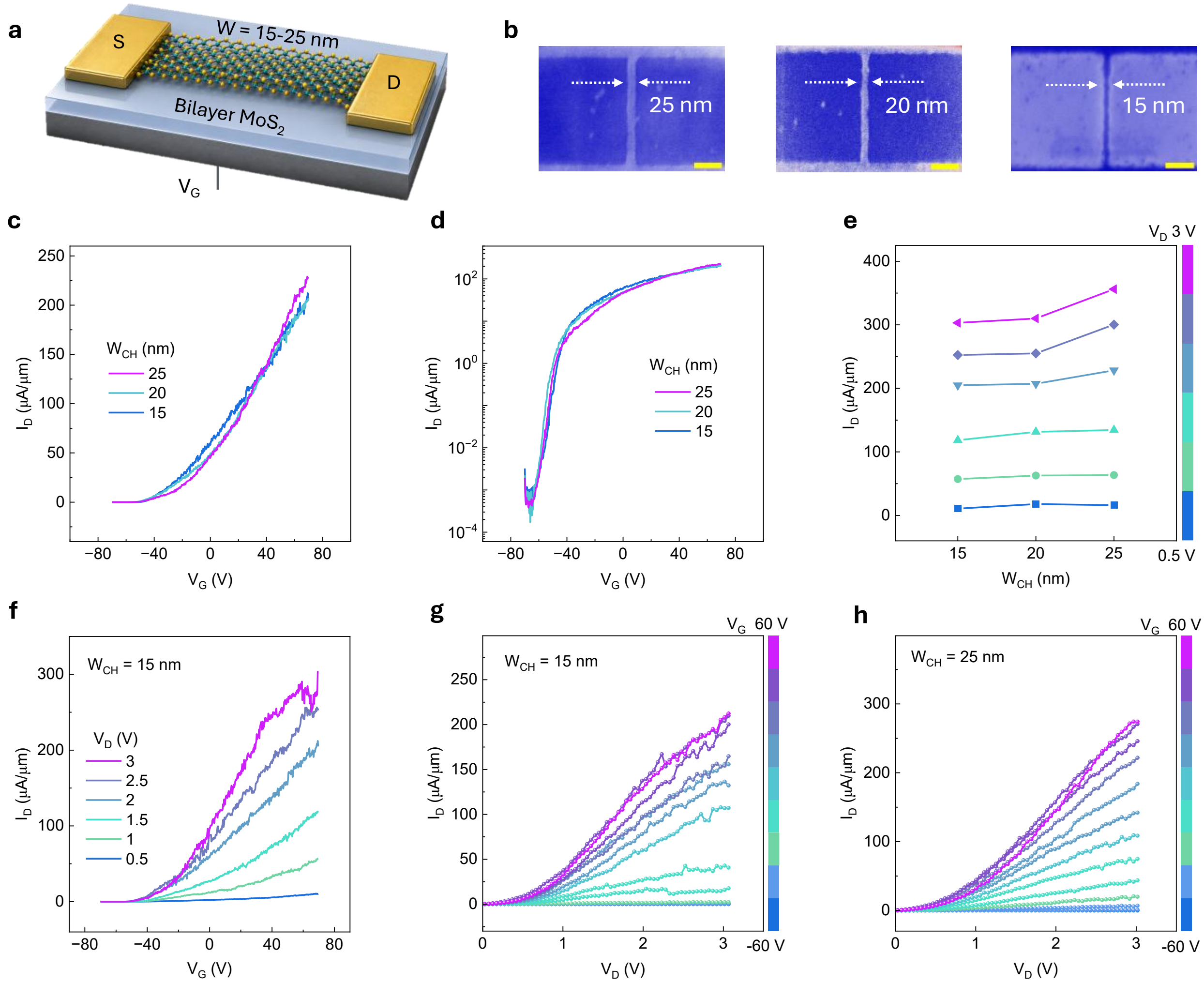

**Figure 4. Ultra-scaled high-performance bilayer $MoS_2$ nanoribbon transistor down to a width of 15 nm. (a)** Schematic of a back-gated bilayer $MoS_2$ nanoribbon transistor with channel widths down to 15 nm; Scale bar is 100 nm. **(b)** False-colour high resolution SEM images of ultranarrow bilayer $MoS_2$ nanoribbons with channel widths 25, 20, and 15 nm (left to right). **(c)** Linear-scale transfer characteristics ($I_D$ vs. $V_G$) for bilayer $MoS_2$ nanoribbon transistors with channel widths $W_{CH}$ = 15, 20, and 25 nm at $V_D$ = 2 V and $L_{CH}$ = 200 nm. **(d)** Corresponding semi-logarithmic transfer characteristics for the same devices, demonstrating n-type transistor behavior with a high on/off current ratio. **(e)** On-current density ($I_D$) as a function of $W_{CH}$ for drain voltages ranging from $V_D$ = 0.5 V to 3 V. **(f)** Transfer characteristics of the $W_{CH}$ = 15 nm, $L_{CH}$ = 200 nm device at multiple drain biases ($V_D$ = 0.5-3 V), showing sustained enhancement of on-current at the narrowest widths. **(g-h)**, Output characteristics $I_D$–$V_D$ for the same devices ($W_{CH}$ = 15 and 25 nm, respectively) at gate voltages $V_G$ from −60 to 60 V (colour scale), demonstrating quasi-linear output behaviour and high current densities in the ultranarrow width regime.

The bias-dependent transfer and output characteristics in Fig. 4f-h provide further insight into charge transport in the saturation regime of ultimate-width scaled nanoribbon transistors. For each nanoribbon width, the transfer curves measured at increasing drain bias show a monotonic increase in on-current with $V_D$, while preserving robust gate modulations over the entire bias range (Fig. 4f and Supplementary Fig. S9). The 25 nm wide channel device shows a maximum on current up to 350 μA/μm whereas 20 and 15 nm wide devices show current saturation around 300 µA/µm with abrupt kinks/noise behaviour at $V_D$ = 3 V. The output characteristics in Fig. 4g-h and Supplementary Fig. S10 corroborate this picture more clearly. For all three widths, $I_D$ exhibits quasi-linear behavior at low drain bias, followed by weakly saturating regime at high $V_D$ as $V_G$ is increased from negative to positive values. The linear low-bias region points to reasonably low contact resistance and an absence of current crowding at the nanoribbon-metal interface. At higher $V_D$, small abrupt kinks in the output curves with gate overdrive suggests that velocity saturation and modest self-heating could be limiting the current drive in the 15 nm width nanoribbon device. The saturation of the on-current at the narrowest widths therefore appear to originate primarily from the increasing contribution of edge-related scattering and quasi-ballistic transport limitations. These observations establish an experimentally accessible ultimate width-scaling boundary for 2DSC nanoribbon transistors.

### Benchmarking of ultra-scaled 2D semiconductor nanoribbon transistors

The comparison of the current densities presented in this work is plotted with earlier 2DSC-based nanoribbon transistors for different thicknesses ranging from atomically thin mono-bilayers to few-nanometer thicknesses, as displayed in Fig. 5a. Earlier studies of 50 nm-wide monolayer $MoS_2$ nanoribbon transistors reported maximum on-current densities of 38 µA/µm[21], comparable to the corresponding 50 nm-wide monolayer devices demonstrated here. However, scaling the channel width down to 30 nm in our devices increases the on-current density to 110 µA/µm, representing an approximately threefold enhancement over 50 nm wider ribbons. Previous multilayer $MoS_2$ nanoribbon transistors with thicknesses of 6-20 nm achieved their highest drive currents only for channel widths exceeding 100 nm, while scaling the ribbon width down to 60 nm, the normalized on-current typically saturated or degraded because transport became increasingly limited by edge roughness, plasma-induced disorder and edge-induced depletion[20,24]. In contrast, the monolayer and bilayer nanoribbons reported here exhibit a non-classical width dependence in which the normalized on-current density increases continuously as the ribbon width is reduced from 80 to 30 nm. Within this enhancement regime, bilayer devices achieve current densities ∼400 $\mu A\ \mu m^{-1}$ at 30 nm widths, substantially exceeding previous publications at comparable dimensions. The shaded regions in Fig. 5a distinguish the enhancement regime from the saturation regime emerging below 30 nm, where further width scaling no longer produces proportional increases in current density down to 15 nm. Furthermore, the nanoribbons reported here also exhibit higher on-current densities than recently reported $MoS_2$ transistors with sub-35 nm channel length and contact lengths but much wider (~800 nm) channel widths[18].

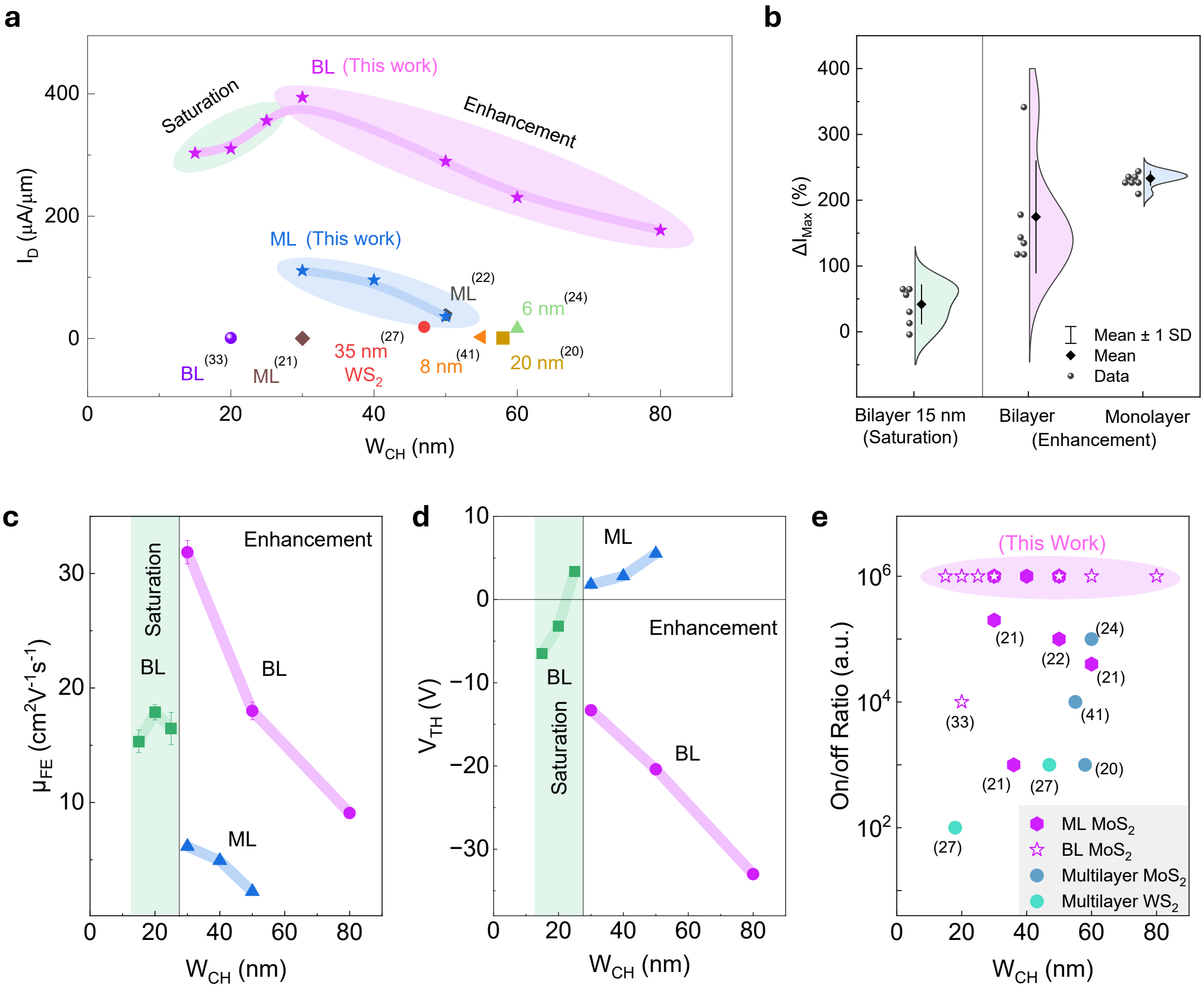

**Figure 5. Benchmarking ultra-width-scaled transport parameters of monolayer and bilayer $MoS_2$ nanoribbon transistors. (a)** Benchmarking of on-current density ($I_D$) as a function of channel width ($W_{CH}$) for monolayer (ML) and bilayer (BL) $MoS_2$ nanoribbon transistors in this work, compared with previous literature reports of 2DSC-based nanoribbon devices with varying thicknesses (colored symbols). The shaded magenta and blue regions define the enhancement regime followed by a saturation regime (shaded green region) in this work demonstrates superior on-current densities compared to prior reports. **(b)** Distribution of the percentage change in maximum on-current density $\Delta I_{\text{max}}$ for devices in the saturation (BL, 15 nm) and enhancement (BL and ML) regimes, showing that narrowing widths down to 15 nm preserves high $I_D$. **(c)** Width dependence of the extrinsic field-effect mobility $\mu_{FE}$ for ML and BL nanoribbons, highlighting mobility enhancement in the 30-80 nm BL devices and a nearly constant mobility in the 15-25 nm saturation regime. **(d)** Corresponding threshold voltage $V_{TH}$ as a function of $W_{CH}$, indicating a strong positive shift with decreasing width in the enhancement regime that evolves into a weakly width-dependent plateau once the nanoribbon becomes narrower. **(e)** Benchmarking of on/off current ratios of nanoribbon transistors with different widths from this work against previous reports on 2DSC-based nanoribbon devices, demonstrating that both the enhancement and saturation regimes maintain on/off ratios above $10^6$.

The distributions of the maximum current enhancement at different drain biases further highlight that the mean percentage enhancement is about 230% for monolayers and 170% for bilayer nanoribbon devices operating in the enhancement regime. The ultranarrow 15 nm nanoribbon devices also retain high drive mean current enhancement of about 45%, as represented in Fig. 5b. The width dependence of the field-effect mobility ($\mu_{\mathrm{FE}}$) shown in Fig. 5c reveals that in the enhancement regime the mobility increases as the channel width decreases for both monolayer and bilayer nanoribbon transistors, in contrast to the previous reports in monolayer $MoS_2$ nanoribbons with 30 nm-wide devices exhibit reduced mobility compared with 60-200 nm-wide ribbons[21]. Bilayer devices retain higher mobilities than their monolayer counterparts across all channel widths. Supplementary Figs. S11 and S12 show the mobility distributions at different drain-source biases in width-scaled monolayer and bilayer $MoS_2$ devices, respectively. The nearly constant mobility observed between 15 and 25 nm reflects that the mobility evolution enters a saturation regime in which further width reduction produces only weak changes in $\mu_{\mathrm{FE}}$. The threshold voltage of monolayer transistors (Fig. 5d) shifts from slightly positive values to zero as the width is reduced from 50 to 30 nm. This contrasts with earlier $MoS_2$ nanoribbon transistors, where reducing the width shifted the $V_{\mathrm{TH}}$ towards positive values, converting depletion-mode into enhancement-mode operation via edge depletion[24]. For bilayer devices, the threshold voltage shifts from negative values to zero as the width is reduced from 80 to 30 nm. This opposite trend between mono- and bilayer devices could arise from interface-induced charge transfer from contacts. However, the convergence of threshold voltages toward near-zero values at narrower widths down to 15 nm suggests enhanced edge electrostatics. The benchmarking plot (Fig. 5e) positions the present monolayer and bilayer $MoS_2$ nanoribbon transistors at the forefront of width-scaled 2DSC devices, achieving on/off ratios exceeding $10^6$ values that surpass those of earlier $MoS_2$ and $WS_2$ nanoribbon transistors reported at comparable channel widths, even those obtained in thin and multilayer channels. Note: While working on this paper, we became aware of two preprints that also report enhanced performance in 2DSC nanoribbons down to 30-35 nm, demonstrating reproducibility of results across different laboratories[44,45]. However, our work goes beyond that, demonstrating high-performance 2DSC nanoribbon transistors in an ultranarrow regime down to channel width of 15 nm.

## Conclusions and outlook

Our results establish a new ultra-narrow width-scaling principle for 2D semiconductors down to 15 nm with improved performance that is distinct from conventional semiconductor physics. By reducing the channel width of monolayer and bilayer $MoS_2$ nanoribbon transistors from 80 to 15 nm, we observe an enhancement in on-current density and saturation at narrower transistors. This directly reverses the degradation trend reported earlier in silicon and 2D semiconductors at comparable dimensions. The improvements in field-effect mobility and threshold-voltage stability reveal an edge-enhanced transport regime and suppressed edge depletion, reflecting an intrinsic electrostatic advantage of atomically thin nanoribbon channels. From a technological perspective, the enhancement in on-current density down to 15 nm wide channel translates directly into circuit-level benefits: a target drive current can be delivered with less total channel width, reducing gate capacitance and dynamic switching energy proportionally, which can position 2DSC nanoribbons as a path toward higher-density, lower-power logic. Combined with the demonstrated compatibility of this top-down lithographic approach, these findings provide an experimental foundation for channel-width scaling for the ultimate scaling of transistor technologies.

## Methodology

**Preparation of $MoS_2$ nanoribbons:** Monolayer and Bilayer $MoS_2$ are mechanically exfoliated from bulk $MoS_2$ (HQ Graphene), onto a 285 nm silicon dioxide ($SiO_2$)/$p^{++}$-Si substrate. The fabrication of $MoS_2$ nanoribbons is carried out using electron beam lithography (Raith EBPG 5200, 100 kV acceleration voltage) and inductively coupled plasma/reactive ion etching (ICP/RIE). The electron beam (e-beam) resist AR-P 6200.13 1:2 dilution in anisole is spin-coated at 6000 rpm resulting in a resist thickness of approximately 50 nm, and then soft baked at 150 °C for 60 seconds. The nanoribbons are defined with EBL with a dose of 240 $\mu C/cm^2$ and are then developed with n-amyl acetate for 80 seconds. The thin nature of the resist combined with a low EBL writing dose reduces scattering of the electrons in the resist. Hence, broadening of the nanoribbon designs is decreased and fabrication of ultra-narrow $MoS_2$ nanoribbons with widths according to design is possible. The exposed areas are etched using ICP/RIE with trifluoromethane ($CHF_3$)/argon (Ar) (50/40 sccm) for 5 seconds. A gentle etching procedure, focusing on ICP and the inert gas Ar for sharp and straight etching profiles and minimal damage to the nanoribbons. During the ICP/RIE, the top layer of the AR-P may get hardened, making it difficult to dissolve the resist. Therefore, an additional step of oxygen ($O_2$) plasma stripping (10 sccm) for 10 seconds is performed. The remaining resist is dissolved in acetone for 24 hours and then cleaned in IPA for 5 minutes.

**Nanoribbon characterization:** The channel length and width of the $MoS_2$ nanoribbons are obtained using the Zeiss Supra 60 VP, EOS Gemini 1, scanning electron microscope (SEM). The etching depth and rate are evaluated with atomic force microscopy (AFM) with tapping mode using the Bruker dimension ICON scanning probe microscope (SPM).

**Raman measurements:** Raman measurements are conducted using a WITec alpha300 R confocal Raman microscope at room temperature. A 532-nm-wavelength laser with a power of 1 mW was used to excite the samples. A diffraction grating (1800 gratings $mm^{-1}$) is used to achieve suitable spectral resolution producing a spot size of approximately 1 μm.

**Device fabrication**: The source and drain contacts are defined using EBL (Raith EBPG 5200, 100 kV acceleration voltage) and then developed with n-amyl acetate for 50 seconds followed by methyl isobutyl ketone (MIBK):IPA 1:1 for 1 minute and 10 seconds. A bilayer MMA EL8/AR-P 6200.13 1:2 e-beam resist stack spin-coated at 6000 rpm is used. The deposition of the top contacts is carried out with electron beam physical vapor deposition (EBPVD), depositing 15 nm nickel (Ni) and 20 nm gold (Au). Finally, the redundant metal is lifted off by submerging in acetone for 30 minutes, followed by cleaning in IPA for 5 minutes.

**Electrical characterization:** Electrical measurements for transistor characteristics are performed with a dual channel source measure unit Keithley 2612B and a single channel source measure unit Keithley 2400. All measurements are performed at room temperature (295 K) under vacuum ($\sim 10^{-4}$ mbar).

## Data Availability

Relevant data supporting the key findings of this study are available within the article and the Supplementary Information file. All raw data generated during the current study are available from the corresponding authors upon request.

## Acknowledgment

Authors acknowledge funding from European Innovation Council (EIC) Graphene Flagship project 2DSPIN-TECH (No. 101135853), 2D TECH VINNOVA competence center (No. 2019-00068) and Areas of Advance (AoA) Nano and Energy programs. Devices were fabricated at the Nanofabrication Laboratory MyFab at Chalmers University of Technology.

## Author Contributions

S.K.M. and S.P.D. conceived the idea and designed the experiments. S.K.M. and A.C. fabricated and characterized the devices and performed the electrical measurements. S.K.M. and S.P.D. analyzed the results and wrote the paper by taking input from all the coauthors. S.P.D supervised the project.